\documentclass{IEEEtran}
\usepackage{stmaryrd}
\usepackage{amsfonts}
\usepackage{graphics} 
\usepackage{subfigure}
\usepackage{epsfig} 
\usepackage{mathptmx} 
\usepackage{times} 
\usepackage{amssymb}  
\usepackage{floatrow}
\usepackage{multirow}
\usepackage{lettrine}
\usepackage{algorithm}
\usepackage{algpseudocode}
\usepackage{color}
\usepackage{booktabs}
\usepackage{graphicx}
\usepackage{makecell}
\usepackage{supertabular}
\usepackage{amsmath} 

\newtheorem{definition}{\indent Definition}

\newtheorem{proposition}{\indent Proposition}
\newtheorem{remark}{Remark}

\IEEEoverridecommandlockouts    

\title{Learning-based Quantum Robust Control: Algorithm, Applications and Experiments}
\author{Daoyi Dong,
Xi Xing, Hailan Ma, Chunlin Chen, Zhixin Liu, Herschel Rabitz
\thanks{This work was supported by the Australian Research
Council's Discovery Projects funding scheme under Project
DP190101566, the National Natural Science Foundation of China
(No. 61828303 and No. 61833010), NSF (CHE-1464569) and ARO (W911NF-16-1-0014).}
\thanks{D. Dong is with the School of Engineering and Information
Technology, University of New South Wales, Canberra, ACT 2600,
Australia, and the Department of Chemistry, Princeton University, Princeton, NJ 08544, USA. (email: daoyidong@gmail.com).}
\thanks{X. Xing is with the Department of Chemistry, Princeton University, Princeton, NJ 08544, USA. (email: xxing@princeton.edu).}
\thanks{H. Ma and C. Chen are with the Department of Control and Systems Engineering, School of Management and Engineering, Nanjing
University, Nanjing 210093, China.(e-mail: clchen@nju.edu.cn).}
\thanks{Z.X. Liu is with the Key Laboratory of Systems and Control, Academy of Mathematics and Systems Science, Chinese Academy of Sciences, Beijing 100190, China (e-mail: lzx@amss.ac.cn).}
\thanks{H. Rabitz is with the Department of Chemistry, Princeton University, Princeton, NJ 08544, USA. (email: hrabitz@princeton.edu).}
}
\begin{document}
\maketitle

\begin{abstract}
Robust control design for quantum systems has been recognized as a key task in quantum information technology, molecular chemistry and atomic physics. In this paper, an improved differential evolution algorithm, referred to as \emph{msMS}\_DE, is proposed to search robust fields for various quantum control problems. In \emph{msMS}\_DE, multiple samples are used for fitness evaluation and a mixed strategy is employed for the mutation operation. In particular, the \emph{msMS}\_DE algorithm is applied to the control problems of (i) open inhomogeneous quantum ensembles and (ii) the consensus goal of a quantum network with uncertainties. Numerical results are presented to demonstrate the excellent performance of the improved machine learning algorithm for these two classes of quantum robust control problems. Furthermore, \emph{msMS}\_DE is experimentally implemented on femtosecond laser control applications to optimize two-photon absorption and control fragmentation of the molecule $\text{CH}_2\text{BrI}$. Experimental results demonstrate excellent performance of \emph{msMS}\_DE in searching for effective femtosecond laser pulses for various tasks.
\end{abstract}

\section{Introduction}\label{sec:Sec1}
Estimating and controlling quantum systems has become a fundamental task in developing quantum technologies, as well as being of basic relevance to many emerging areas in atomic physics, chemistry, and quantum information science \cite{Dong and Petersen 2010}-\cite{Pan}. Some control methods, such as optimal control theory \cite{Dong and Petersen 2010}, learning control algorithms \cite{Rabitz et al 2000} and Lyapunov control approaches \cite{c10}, \cite{Kuang et al 2017}, have been developed for manipulating quantum systems. Among these methods, learning control is a powerful approach for many complex quantum control tasks and has achieved great success in laser control of molecules and other applications since the method was presented in the seminal paper \cite{Judson and Rabitz 1992}.  Many quantum learning control problems can be formulated as an optimization problem and a learning algorithm is employed to search for an optimal control field for achieving desired performance. Gradient algorithms have been demonstrated to be a good candidate for numerically finding an optimal field due to their high efficiency \cite{Khaneja et al 2005}. In many practical applications, the gradient information may not be easy to obtain and some complex quantum control problems may have local optima if certain conditions are not satisfied \cite{Rabitz et al 2004}, \cite{Chakrabarti and Rabitz 2007}. For these situations, it is necessary to employ stochastic search algorithms to find a good control field. The genetic algorithm (GA) has been widely used in the area of quantum control and has achieved success in learning control of molecular systems \cite{Rabitz et al 2000}. In this paper, we focus on robust control problems of quantum systems and explore the use of differential evolution (DE) \cite{Storn and Price 1997} to search for robust control fields.

DE is a competitive form of evolutionary computation and has shown great performance for many complex optimization problems \cite{Das and Suganthan 2011}. Recently, it has also been used for solving quantum control problems \cite{Zahedinejad et al 2015}, \cite{Ma 2015SMC}, \cite{Ma2017CTT}. For example, Zahedinejad \emph{et al.} \cite{Zahedinejad et al 2015}, \cite{Zahedinejad et al 2016} proposed a subspace-selective self-adaptive differential evolution (SUSSADE) algorithm to achieve a high-fidelity single-shot Toffoli gate and single-shot three-qubit gates. DE methods have been employed to achieve a desired state transfer by designing optimal control fields for an open quantum ensemble \cite{Sun et al 2015}. Zahedinegad \emph{et al.} \cite{Zahedinejad et al 2014} investigated several promising evolution algorithms and found DE outperformed GA and particle swarm optimization for hard quantum control problems. In this paper, we employ DE algorithms to solve several classes of quantum robust control problems. The robustness of quantum systems is an essential requirement for the development of practical quantum technology, and robust control methods could provide enhanced robust performance for quantum systems \cite{James et al 2008}, \cite{Dong 2012Automatica}, \cite{Wang-Dong-TAC}, \cite{WuR-PRA-2019}. In particular, we propose an \emph{msMS}\_DE (\textbf{m}ultiple-\textbf{s}amples and \textbf{M}ixed-\textbf{S}trategy DE) algorithm where a mixed strategy and an average performance with multiple samples are employed. The \emph{msMS}\_DE is used for three classes of quantum robust control problems: control of inhomogeneous open quantum ensembles, consensus of quantum networks with uncertainties and experimental fragmentation control using femtosecond laser pulses \cite{Xing NJP}.

A quantum ensemble consists of many single quantum systems (e.g., atoms, molecules or spin systems) and every individual system is referred to as a member of the ensemble \cite{Chen et al 2014}.
Here, an ensemble is considered in the sense of the individual systems with slightly different parameter values, rather than in the sense of mixed states.
Such a case is called an inhomogeneous quantum ensemble \cite{Chen et al 2014}, \cite{Li and Khaneja 2006}. Inhomogeneous quantum ensembles have wide applications in fields ranging from magnetic-resonance imaging to quantum communication \cite{Li et al 2011}, \cite{Duan et 2001}. Hence, it is highly desirable to design control laws that can steer an inhomogeneous ensemble from a given initial state to a target state when variations exist in the system parameters.
Some results have been presented for controllability analysis and control design of inhomogeneous quantum ensembles. For example, Li and co-workers \cite{Li and Khaneja 2006}, \cite{Li et al 2011} presented a series of results to analyze controllability and design optimal control laws for inhomogeneous spin ensembles. A Lyapunov control design approach has been proposed to asymptotically stabilize a spin ensemble around a uniform state of spin $+1/2$ or $-1/2$ \cite{Rouchon-inhomogeneous}.  Chen \emph{et al.} \cite{Chen et al 2014} presented a sampling-based learning control method to achieved high fidelity control of inhomogeneous quantum ensembles. In these results, decoherence and dissipation were usually not considered. The existence of decoherence and dissipation may irreversibly lead a quantum ensemble to becoming an open system \cite{Breuer and Petruccione 2002}, and the manipulation of inhomogeneous open quantum ensembles becomes more challenging than without considering decoherence. For an inhomogeneous quantum ensemble, we cannot employ different control fields to control individual members. A practical solution is to find a robust control field that can drive all of the members in the ensemble into a given target state. In this paper, we employ an \emph{msMS}\_DE algorithm to search for such robust control fields for inhomogeneous open quantum ensembles aiming at achieving enhanced control performance.

Another problem under consideration is to drive a quantum network into a consensus state even in the presence of uncertainties. Achieving consensus is one of primary objectives in distributed coordination and control of classical (non-quantum) networked systems \cite{Mesbahi and Egerstedt 2010}. Consensus usually means that all of the nodes in a network hold the same state. Recent development in quantum technology has made it significant and feasible to analyze quantum networks where each node (agent) represents a quantum system such as a photon, an electron, a spin system or a superconducting quantum bit (qubit) \cite{Mazzarella et al 2015}, \cite{Shi et al 2014}. Consensus of quantum networks may have potential applications in promising quantum communication networks, distributed quantum computation and one-way quantum computation \cite{Mazzarella et al 2015}, \cite{Shi et al 2014}. Since the nodes in a quantum network are quantum systems, some unique characteristics such as quantum entanglement and measurement backaction, different from classical multi-agent systems \cite{Chenglong2014}, \cite{Liu2017}, should be carefully considered, and the analysis and control of quantum networks raise new challenges. Some results have been presented for the consensus problem of quantum networks. For example, Sepulchre \emph{et al.} \cite{Sepulchre et al 2010} generalized consensus algorithms to noncommutative spaces and analyzed the asymptotic convergence to the consensus state of a fully mixed state. Ticozzi and co-workers \cite{Mazzarella et al 2015}, \cite{Ticozzi 2}, \cite{Ticozzi 3} presented a series of results on consensus of quantum networks including several different definitions for quantum consensus, quantum gossip algorithms, and quantum consensus results within a group-theoretic framework. Shi \emph{et al.} \cite{Shi et al 2014} presented a systematic investigation on consensus of quantum networks with continuous-time dynamics within the framework of graph theory. In this paper, we consider the basic problem of finding a robust control law to steer a quantum network to a reduced state consensus (as defined in \cite{Mazzarella et al 2015}) and do not consider the distributed solutions to achieving quantum consensus. Achieving quantum consensus states has significant applications in one-way quantum computation, initialization of quantum networks and distributed quantum computation. In particular, we employ the proposed \emph{msMS}\_DE for driving a superconducting qubit network with uncertainties into a reduced state consensus.

The third problem that we consider illustrating the diverse capability of the proposed algorithm is fragmentation control of molecules using femtosecond laser pulses. Femtosecond (fs) ($1 \text{fs}=10^{-15}$ second) lasers \cite{Xing NJP} have found wide applications in controlling molecular dynamics because of their short pulse duration, which is comparable to the time scales of electronic and nuclear motions of a molecule. The temporal structures of a femtosecond pulse could be manipulated by pulse shaping techniques \cite{Xing NJP}, which is typically achieved by modulating the phase and/or amplitude of the laser frequency components, spatially separated and manipulated, with a computer programmable spatial light modulator (SLM) before recombination into a ``shaped pulse". In quantum control experiments, a practical approach is to use closed-loop learning control \cite{Rabitz et al 2000} to find an optimal field that can steer the quantum system towards the desired outcome, which has achieved much success.  An evolutionary algorithm (e.g., GA) is often employed to assist the search for an optimal pulse. To the best of our knowledge, there have not been any experimental results reported where DE algorithms are used on femtosecond laser control experiments. There are few quantum control experiments using femtosecond laser pulses that investigated robustness to variations in the control. In this work, we employ the \emph{msMS}\_DE algorithm to an experimental quantum control problem, where the goal is to identify a robust solution (shaped fs laser pulse) that can maximize the $\text{CH}_2\text{Br}^{+}/\text{CH}_2\text{I}^{+}$ product ratio from the fragmentation of the $\text{CH}_2\text{BrI}$ molecule in a time-of-flight mass spectrometry (TOF-MS). The \emph{msMS}\_DE algorithm is also used in a separate experiment to identify the transform limited (TL) pulse via optimizing the two photon absorption (TPA) signal, which is carried out prior to the fragmentation experiment.

This paper focuses on employing machine learning to design robust fields for three classes of quantum control problems. The main contributions of this paper are summarized as follows:
\begin{itemize}
  \item Motivated by solving three classes of quantum robust control problems, an improved DE algorithm of \emph{msMS}\_DE is proposed where an average fitness function of multiple samples is utilized and a mixed strategy of mutation is employed.
  \item The control problem of inhomogeneous open quantum ensembles is investigated and the \emph{msMS}\_DE algorithm is used to learn robust control fields for inhomogeneous open quantum ensembles. Numerical results show that the control fields learned by \emph{msMS}\_DE usually have improved robust performance compared with those learned by basic DE and GA.
  \item The task of driving a quantum network with uncertainties to a consensus state is investigated and robust control fields can be found by employing the \emph{msMS}\_DE algorithm. Numerical results demonstrate that \emph{msMS}\_DE has excellent performance in searching for robust control fields for achieving consensus of quantum networks.
  \item Several experiments are implemented on femtosecond laser control systems where the \emph{msMS}\_DE algorithm is employed to find effective laser pulses for generating an excellent TPA signal and achieving good fragmentation control of molecules. These experiments present the first tests of DE for femtosecond laser quantum control as well as realize the sampling-based learning control method \cite{Chen et al 2014}, \cite{Dong et al 2015}, \cite{Wu et al 2017}.
\end{itemize}

The paper is organized as follows. Section \ref{sec:Sec2} formulates three classes of quantum robust control problems under consideration. Section \ref{sec:Sec3} presents a systematic description of the \emph{msMS}\_DE algorithm. Numerical results on quantum ensemble control and quantum network consensus are provided in Section \ref{sec:Sec4}. In Section \ref{sec:Sec5}, we present experimental results on femtosecond laser control. Concluding remarks are given in Section \ref{sec:Sec7}.

\section{Three classes of quantum control problems}\label{sec:Sec2}
\subsection{Control of inhomogeneous open quantum ensembles}\label{subsec:Sec2-2}
We consider an inhomogeneous quantum ensemble where the parameters describing the system dynamics of the ensemble members could have variations. An example is that a spin ensemble in nuclear magnetic-resonance (NMR) may encounter quite large dispersion in the strength of the applied radio frequency field (rf inhomogeneity) and there also exist variations in the natural frequencies of these spins (Larmor dispersion) \cite{Li and Khaneja 2006}. Several methods have been proposed for control law design of inhomogeneous quantum ensembles when dissipation and decoherence were not considered \cite{Chen et al 2014}, \cite{Li et al 2011}.

For a practical quantum ensemble, each member in the ensemble should be dealt as an open quantum system. The state of an open quantum system can be described by a positive Hermitian density operator $\rho$ with the constraint $\text{tr}(\rho)=1$. A Markovian master equation for $\rho(t)$ can be used to describe the dynamics of an open quantum system interacting with its environment when we assume a short environmental correlation time and permit to neglect memory effects \cite{Breuer and Petruccione 2002}. The Markovian master equation in the Lindblad form is described as \cite{Wiseman and Milburn 2010}, \cite{Lindblad 1976}
      \begin{equation}\label{eq:markovian equations}
         \dot{\rho}(t)=-\frac{\textit{i}}{\hbar}[H(t),\rho(t)]+\sum_{k}\gamma_k\mathcal{D}[L_k]\rho(t),
      \end{equation}
where $i=\sqrt{-1}$, $H(t)$ is the system Hamiltonian, $\hbar$ is Planck's constant (hereafter we will set $\hbar=1$), the non-negative coefficients $\gamma_k$ specify the relevant relaxation rates, $L_k$ are appropriate Lindblad operators and $$\mathcal{D}[L_k]\rho=(L_k \rho L_k^{\dagger}-\frac{1}{2} L_k^{\dagger} L_k \rho-\frac{1}{2}\rho L_k^{\dagger} L_k ).$$

For an inhomogeneous open quantum ensemble, the Hamiltonian can be described in the form of
      \begin{equation}\label{eq:system Hamiltonian}
     H_{\varTheta}(t)=g_0(\theta_0)H_0+\sum_{j=1}^{M}g_j(\theta_j)u_j(t)H_j,
     \end{equation}
where we assume that $M$ control Hamiltonians are used. Let $\varTheta=(\theta_0,\theta_1,...,\theta_M)$ and the functions $g_j(\theta_j)$ $(j=0,1,\ldots,M)$ characterize possible inhomogeneities. For example, $g_0(\theta_0)$ corresponds to inhomogeneity in the free Hamiltonian (e.g., due to chemical shift in NMR). $g_j(\theta_j)$ $(j=1,...,M)$ can characterize imprecise parameters in the dipole approximation or possible multiplicative noises in the control fields. We assume that $g_j(\theta_j)$ $(j=0, 1,...,M)$ are continuous functions of $\theta_j$ and the parameters $\theta_j$ could be time-dependent and $\theta_j \in[1-E_j,1+E_j]$. For simplicity, we assume that $g_j(\theta_j)=\theta_j$ in this class of quantum control problems, the nominal value of $\theta_j$ is 1 and $E_0=... = E_j=...= E_M = E $.

For an open quantum system in (1), we may define a coherent vector as $\textbf{y}:=({\rm tr}(U_l \rho),{\rm tr}(U_2 \rho),...,{\rm tr}(U_m \rho))^\top$, where $\textit{i}U_1$, $\textit{i}U_2$, ... $\textit{i}U_m$ $(m=n^2-1)$ are orthogonal generators of the special unitary group $\mathbf{su}(\textit{n})$ with degree $n$ . Its density operator can be written as:
      \begin{equation}\label{eq:density_operator}
            \rho=\frac{I}{n}+\frac{1}{2}\sum_{l=1}^{m}y_l U_l.
      \end{equation}
Substituting (\ref{eq:density_operator}) into (\ref{eq:markovian equations}), the evolution of the coherent vector $\textbf{y}$ can be described as:
      \begin{equation}\label{eq:evolution of coherent vectors}
      \dot{\textbf{y}}=(\mathcal{L}_{H_0}+\mathcal{L}_D)\textbf{y}+\sum_{j=1}^{M}u_j\mathcal{L}_{H_j}\textbf{y}+l_0,
      \end{equation}
where the superoperators $\mathcal{L}_{H_0}$, $\mathcal{L}_D$, $\mathcal{L}_{H_j}$ $(j=1,2,...,M)$ and the term $l_0$ are explained in detail in \cite{Yang et al 2013}, \cite{Kimura 2003}. We choose the objective function $J(u)$ to be maximized as follows \cite{Yang et al 2013}:
     \begin{equation}\label{eq:objective functions}
     J(u)=1-\frac{n}{8(n-1)}\parallel\textbf{y}_f -\textbf{y}(T)\parallel^2,
      \end{equation}
where $\parallel x \parallel^2 =x^Tx $ is a vector norm and it is clear that $J(u)\in [0,1]$. And, $\textbf{y}_f$ and $\textbf{y}(T)$ are the target state and the final state of the quantum system in terms of coherent vector, respectively.

The control of an inhomogeneous open quantum ensemble can be formulated as:
\begin{equation}\label{eq:formulation of open ensemble}
\begin{split}
\displaystyle  \ \ \  & \max_u \
J(u):=\max_{u} \mathbb{E}[J_\varTheta(u)]\\
\text{s.t.} \\
&\left \{
\begin{split}
&\dot{\textbf{y}}_\varTheta(t)=(\theta_{0}\mathcal{L}_{H_0}+\mathcal{L}_D+\theta_j\sum_{j=1}^{M}u_j(t)\mathcal{L}_{H_j})\textbf{y}_\varTheta(t)+l_0\\
&\textbf{y}_\varTheta(0)=\textbf{y}_0,\quad t \in [0, T]\\
&\theta_j\in[1-E,1+E],\quad j=0,1,...M\\
\end{split}\right.\\
\end{split}
\end{equation}
where $J_{\varTheta}(u)$ is the objective function for given $\varTheta$ and $\mathbb{E}[J_{\varTheta}(u)]$ denotes the average performance of $J(u)$ over the parameter inhomogeneities $\varTheta$.

\subsection{Consensus in quantum networks}\label{subsec:Sec2-4}
Achieving quantum consensus is a primary objective in the investigation of quantum networks. Existing results presented some distributed solutions to quantum consensus problems. For example, Mazzarella \emph{et al.} \cite{Mazzarella et al 2015} proposed a quantum gossip iteration algorithm where discrete-time quantum swapping operations between two arbitrary nodes are used to make a quantum network achieving consensus. A graphical method has been developed in \cite{Shi et al 2014} to build the connection between quantum consensus and its classical counterpart, and asymptotic convergence results on achieving a consensus state have been presented for a class of quantum networks with continuous-time Markovian dynamics. Here, we do not intend to develop a distributed algorithm for quantum consensus. In contrast, we consider how to design a robust control field to drive a quantum network from an initial state into a consensus state with high fidelity when uncertainties or inaccuracies may exist in the system dynamics. We consider the type of Reduced State Consensus (RSC) that was defined in \cite{Mazzarella et al 2015}. We denote $\mathcal{H}$ as a Hilbert space, $A\otimes B$ is the tensor product of $A$ and $B$, and $|a\rangle$ is a vector in $\mathcal{H}$, i.e., $|a\rangle\in \mathcal{H}$ \cite{Nielsen and Chuang 2000}. In order to present the definition of reduced state consensus, we need to use the concept of partial trace defined as follows.

\begin{definition}[Partial trace]\cite{Nielsen and Chuang 2000}\label{def:partial trace}
Let $\mathcal{H}_A$ and $\mathcal{H}_B$ be the state spaces of two quantum systems A and B, respectively. Their composite system is described as a density operator $\rho^{AB}$. The partial trace over system B, denoted as $\textup{Tr}_{\mathcal{H}_B}$ is given in the following form
\begin{equation}
\textup{Tr}_{\mathcal{H}_B}(|a_1\rangle \langle a_2 |\otimes |b_1\rangle \langle b_2|)=|a_1\rangle \langle a_2|\textup{Tr}(|b_1\rangle \langle b_2|),
\end{equation}
where the vectors $|a_1\rangle, |a_2\rangle$ $\in \mathcal{H}_A$, and the vectors $|b_1\rangle, |b_2\rangle$ $\in \mathcal{H}_B$.
When the composite system is in the state $\rho^{AB}$, the reduced density operator for system $A$ is defined as $\rho_A=\textup{Tr}_{\mathcal{H}_B}(\rho^{AB})$ and the reduced density operator for system $B$ is defined as $\rho_B=\textup{Tr}_{\mathcal{H}_A}(\rho^{AB})$.
\end{definition}

The RSC for a quantum network can be defined as follows.
\begin{definition}[Reduced State Consensus]\cite{Mazzarella et al 2015}\label{def:RSC}
A quantum network consisting of $m$ nodes with the state $\bar{\rho}$ is in a RSC if
\begin{equation}
\nonumber
\bar{\rho}_1=\bar{\rho}_2=...=\bar{\rho}_m,
\end{equation}
\end{definition}
where $\bar{\rho}_j=\text{Tr}_{\otimes_{k\neq j} \mathcal{H}_{k}}(\bar{\rho})$ $(j=1,2,...,m)$ is defined as the reduced density operator for node $j$ and can be calculated according to Definition \ref{def:partial trace}.

We aim to steer a quantum network into a consensus state in Definition \ref{def:RSC}. In practical applications, the existence of noise (including intrinsic and extrinsic), inaccuracies (e.g., variation in the coupling between nodes) and fluctuations (e.g., fluctuations in control fields) in quantum networks is unavoidable. We assume that the Hamiltonian with uncertainties can be written as
\begin{equation}
H_{\varTheta}(t)=\theta_0 H_0+\sum_{j=1}^{M}\theta_j u_j(t)H_j.
\end{equation}
The problem can be formulated as follows:
\begin{equation}\label{eq:formulation_network}
\begin{split}
\displaystyle  \ \ \  & \max_u \
J(u):=\max_{u} \mathbb{E}[J_\varTheta(u, \bar{\rho})]\\
\text{s.t.} \\
&\left \{
\begin{split}
&\dot{\rho}(t)=-i\left[\theta_0 H_0+\sum_{j=1}^{M} \theta_j u_j(t)H_j, \rho(t)\right] \\
&\ \rho(0)=\rho^{0},\quad t \in [0, T]\\
&\theta_j\in[1-E,1+E],\quad j=0,1,2,...M\\
\end{split}\right.\\
\end{split}
\end{equation}
where $J_{\varTheta}(u, \bar{\rho})$ is the objective function for given $\varTheta$ and $\bar{\rho}$, $\mathbb{E}[J_{\varTheta}(u, \bar{\rho})]$ denotes the average performance function with respect to the parameter variations $\varTheta$, the target consensus state is $\bar{\rho}$, and $E \in[0,1]$ is the bound of the parameter uncertainties.

\subsection{Femtosecond laser quantum control}

We consider the experimental control of molecular fragmentation using shaped femtosecond laser pulses. Here, $\text{CH}_{2}\text{BrI}$ is chosen as the target molecule.  As a family member of halomethane molecules, whose dissociative products play a central role in ozone depletion, $\text{CH}_{2}\text{BrI}$ has attracted wide attention because of its importance in environmental chemistry. In addition, it is one of the simplest prototype molecules containing different bonds, a stronger C-Br bond and a weaker C-I bond, which is ideal for the study of controlling selective bond-breaking. Under strong femtosecond laser pulses, $\text{CH}_{2}\text{BrI}$ molecules will undergo ionization and dissociation, and their charged products can be separated and detected with a TOF-MS. In particular, we choose to optimize the photoproduct ratio of $\text{CH}_2\text{Br}^{+}/\text{CH}_2\text{I}^{+}$ as our control objective, which corresponds to breaking the weak C-I bond versus the strong C-Br bond.  We apply closed-loop learning control, using the proposed \emph{msMS}\_DE algorithm, to search for a robust ultrafast laser pulse that maximizes this ratio.

In closed-loop learning control, the learning process can be conceptually expressed as follows. First, one applies trial input pulses to the molecules subject to control and observes the results. Second, a learning algorithm suggests better control inputs based on the prior experiments. Third, one applies ``better" control inputs to new molecules \cite{Rabitz et al 2000}. This approach has been employed to explore the quantum control landscape \cite{Rabitz et al 2004} to find the optimal control strategy where the control performance function $J(u)$ reaches its maximum.  In order to achieve good performance, we first need to identify a reference phase mask on the SLM that give shortest transform limited (TL) pulse, which can be obtained from optimizing the signal of TPA. Optimal pulses with reference to the TL pulse are not subject to the variations of the reference pulse, and are more meaningful to undergo further analysis and comparison. We first use the proposed \emph{msMS}\_DE algorithm to search for a good control to obtain high TPA signal. Then we apply the same algorithm to search for a good control for the fragment ratio of $\text{CH}_2\text{Br}^{+}/\text{CH}_2\text{I}^{+}$.  The consideration of robustness with multiple samples (MS) in DE would also ensure good transferability of the experimental results or photonic reagents \cite{Tibbetts JCP} to another laboratory. That is, an optimal pulse identified from one laser system would also perform well (if not optimal) when transferred to another system despite the minor differences or uncertainties in the control parameters (i.e., the spectral phases on the SLM) and even in the second laser system setup.

\section{\emph{msMS}\_DE algorithm for quantum robust control}\label{sec:Sec3}
DE was first proposed in 1990s \cite{Storn and Price 1997}, \cite{Storn and Price 1995} and has many variants. DE algorithms have been used in wide ranging applications in diverse areas of science and engineering \cite{Das and Suganthan 2011}, \cite{Sarker et al 2016}. The conventional DE algorithm is briefly introduced in the Appendix. Here, we propose an improved DE algorithm, \emph{msMS}\_DE, to solve quantum robust control problems.

In DE, we need to choose appropriate mutation-trial strategies and parameter settings to achieve the success of the algorithm \cite{Qin et al 2009}, \cite{Neri and Tirronen 2010}, \cite{Becerra and Coello 2006}. Several variants of DE utilizing the idea of mixed strategies such as SaDE \cite{Qin et al 2009} and EPDE \cite{Mallipeddi et 2011} have been proposed and exhibited good performance.
Our numerical results show that DE with a single strategy might be adequate for easy problems while DE variants with mixed strategies might be a promising candidate for quantum control problems with multimodal landscapes. Existing results of sampling-based learning control \cite{Chen et al 2014}, \cite{Dong et al 2015}, \cite{quantum classification} have shown that the employment of an average objective function with multiple samples can provide improved performance for quantum robust control problems. Inspired by these observations, we adopt a mixed strategy and an average performance of multiple samples to present an improved DE algorithm (i.e., \emph{msMS}\_DE) for the quantum robust control problems outlined in Section \ref{sec:Sec2}.

We first choose one mutation scheme from a pool of strategy candidates where several mutation schemes with effective yet diverse characteristics are equally distributed. Then, we implement a binomial crossover operation on the corresponding mutant vector to generate the trial vector. Note that we assign various values of $F$ and $CR$ for each individual during the current generation to increase the diversity of the population. To construct the candidate pool, we investigate several commonly used mutation strategies
 \cite{Becerra and Coello 2006} and select four with distinct capabilities at different stages of evolution as follows:

DE/rand/1:
\begin{equation}\label{eq:strategy 1}
    V_i=X_{r_1}+F\cdot(X_{r_2}-X_{r_3}).
\end{equation}

DE/rand to best/2:
\begin{equation}\label{eq:strategy 2}
    V_i=X_i+F\cdot(X_{best}-X_i)+F\cdot(X_{r_1}-X_{r_2})+F\cdot(X_{r_3}-X_{r_4}).
\end{equation}

DE/rand/2:
\begin{equation}\label{eq:strategy 3}
    V_i=X_{r_1}+F\cdot(X_{r_2}-X_{r_3})+F\cdot(X_{r_4}-X_{r_5}).
\end{equation}

DE/current-to-rand/1:
\begin{equation}\label{eq:strategy 4}
    V_i=X_i+K\cdot(X_{r_1}-X_i)+F\cdot(X_{r_2}-X_{r_3}).
\end{equation}
The indices $r_1, r_2, r_3, r_4$ and $r_5$ are mutually exclusive integers randomly chosen from the range $[1,NP]$ and all of them are different from the index $i$. $X_{best}$ is the best individual vector (i.e., the lowest objective function value for a minimization problem) in the population. In the strategy DE/current-to-rand/1, we set the control parameter $K$ as $K=0.5$ to eliminate one additional parameter.
As for the crossover operation, the first three mutation schemes are combined with a binomial crossover operation, while the fourth scheme directly generates trial vectors without crossover.

In the proposed \emph{msMS}\_DE algorithm, we use a normal distribution $\text{N}(0.5,0.3)$ with mean value 0.5 and standard deviation 0.3 to approximate the parameter $F$. It can be verified that values of $F$ fall into the range $[-0.4,1.4]$ with probability of 99.7\% which helps maintain both exploration (with larger $F$ values) and exploitation (with small $F$ values). Also, we let $CR$ obey a normal distribution denoted by N(0.5,0.1) and the small standard deviation 0.1 is enough to guarantee that most values of $CR$ lie in $[0,1]$ \cite{Qin et al 2009}. Consequently, a set of $F$ and $CR$ values are randomly sampled from a normal distribution (denoted by Normrnd) and applied to each target vector in the current population.
We may obtain some extraordinary values far from [-0.4,1.4] for the scale factor $F$ and we usually accept them to increase diversity. While the crossover rate has probabilistic meaning for the chance of survival, we should abandon those falling outside $[0,1]$, and generate another valid parameter by $CR=\text{N}(0.5,0.1)$ to guarantee the practical meaning of crossover.

The \emph{msMS}\_DE method is proposed for three classes of quantum control tasks. In order to design appropriate control laws to achieve good robustness performance, we integrate the idea of sampling-based learning control \cite{Chen et al 2014} into the \emph{msMS}\_DE algorithm. To begin with, we prepare $N$ samples $\varTheta_{k}=(\theta_0,\theta_1,...,\theta_M)$ ($k=1, 2, \dots, N$) with different values of the uncertain parameters. We compute the fitness values of these sample vectors . Then, we evaluate the average fitness value $\bar{f}$ for these samples, and $\bar{f}$ is defined as follows
$$\bar{f}(U_{i,G})=\frac{1}{N}\sum_{k=1}^{N}f(U_{i,G},\varTheta_k),$$
$$\bar{f}(X_{i,G})=\frac{1}{N}\sum_{k=1}^{N}f(X_{i,G},\varTheta_k).$$
The \emph{msMS}\_DE algorithm is outlined in Algorithm \ref{Algorithm EMSDE}.

\begin{algorithm}
\caption{Algorithmic description of \emph{msMS}\_DE}\label{Algorithm EMSDE}
\begin{algorithmic}[1]
\State Set the generation number $G=0$
\For {$i=1$ to $NP$}  
\For {$j=1$ to $D$}
\State  $ x_{i,G}^j=x_{\min}^j+\textup{rand}(0,1)\cdot(x_{\max}^j-x_{\min}^j)$
\EndFor
\EndFor

\State {Initialize fitness $f(X_{i,G})$ and evaluate vector by $f$}
\State {Mark the best vector with maximum $f$ as $X_{\text{best},G}$}

\Repeat {\ (for each generation $G=1,2,\ldots,G_{\max}$)}
\Repeat {\ (for each vector $X_i$, $i=1,2,\ldots,NP$)}

\State Set parameter $F_{i,G}=\text{Normrnd}(0.5,0.3)$
\State Randomly generate a real number pp$\in[0,1]$,

\If  {$pp>0$ and $pp \leq 0.25$} flag=1
\State  {\scalebox{0.8}{$V_{i,G}=X_{r_1,G}+F_{i,G}\cdot(X_{r_2,G}-X_{r_3,G})$}}
\EndIf

\If  {$pp>0.25$ and $pp \leq 0.5$} flag=2
\State {\scalebox{0.8}{$V_{i,G}=X_{i,G}+F_{i,G}\cdot(X_{\text{best},G}-X_{i,G})+F_{i,G}\cdot(X_{r_1,G}-X_{r_2,G})$}}\\
{\scalebox{0.8}{$+F_{i,G}\cdot(X_{r_3,G}-X_{r_4,G})$}}
\EndIf

\If  {$pp>0.5$ and $pp \leq 0.75$} flag=3
\State  {\scalebox{0.8}{$V_{i,G}=X_{r_1,G}+F_{i,G}\cdot(X_{r_2,G}-X_{r_3,G})+F_{i,G}\cdot(X_{r_4,G}-X_{r_5,G})$}}
\EndIf

\If  {$pp>0.75$ and $pp \leq 1$} flag=4
\State  {\scalebox{0.8}{$V_{i,G}=X_{i,G}+K\cdot(X_{r_1,G}-X_{i,G})+F_{i,G}\cdot(X_{r_2,G}-X_{r_3,G})$}}
\EndIf

\For {$i=1$ to $D$}
\If  {$v_{i,G}^j>x_{max}^j$ or $v_{i,G}^j<x_{min}^j$}
\State $v_{i,G}^j=x_{min}^j+\textup{rand}(0,1)\cdot(x_{max}^j-x_{min}^j)$
\EndIf
\EndFor
\State Set parameter $CR_{i,G}=\text{Normrnd}(0.5,0.1)$
\While {$CR_{i,G}<0$ or $CR_{i,G}>1$}
\State $CR_{i,G}=\text{Normrnd}(0.5,0.1)$\
\EndWhile

\If  {flag=1 or flag=2 or flag=3}
\For {$j=1$ to $D$}
\State  {\scalebox{0.8}{$u_{i,G}^j=v_{i,G}^j$, if ($\text{rand}[0,1] \leq CR_{i,G}$ or $j=j_{\text{rand}}$)}}
\State  {\scalebox{0.8}{$u_{i,G}^j=x_{i,G}^j$, otherwise}}
\EndFor
\EndIf

\If  {flag=4} $U_{i,G}=V_{i,G}$
\EndIf
\For {each sample $\varTheta_k$, $(k=1,2,...,N)$ }
\State {evaluate the fitness function $f(U_{i,G},\varTheta_k)$} \label{algo:compute fitness}
\EndFor

\State {Compute $\bar{f}(U_{i,G})=\frac{1}{N}\sum_{k=1}^{N}f(U_{i,G},\varTheta_k)$} \label{algo:compute average}

\If  {$\bar{f}(U_{i,G}) \geq \bar{f}(X_{i,G})$} \label{algo:renew begin}
\State  $X_{i,G+1} \leftarrow U_{i,G}$,\quad $\bar{f}(X_{i,G+1}) \leftarrow \bar{f}(U_{i,G})$.
\EndIf \label{algo:renew end}

\State {Renew the best vector $X_{\text{best},G}$ and $i \leftarrow i+1$}
\Until {\ $i=NP$}

\State  $G \leftarrow G+1$
\Until {\ $G=G_{\max}$}
\end{algorithmic}
\end{algorithm}

\begin{remark}\label{rem:inilization}
In simulations, after we obtain the nominal value of an individual, we may generate the other samples by perturbing the nominal value and then calculate the average fitness function. In experiments, we need to measure the fitness of each sample, and then calculate the average fitness. Usually, a larger number of samples $N$ may lead to better robustness performance \cite{Chen et al 2014}. However, the computational or experimental time will significantly increase with the size of $N$. In this paper, we use three samples for each uncertain parameter to reduce computational and experimental time.
\end{remark}


\begin{remark}\label{rem:terminal condition}
In the proposed \emph{msMS}\_DE algorithm, we preset a maximum generation $G_{\max}$ as the termination criterion. During the implementation of the algorithm, the population evolves until the learning process reaches $G=G_{\max}$. In numerical examples, we let $G_{\max}=50000$. In experimental examples, we choose $G_{\max}=150$.
\end{remark}

\section{Numerical results for ensemble control and quantum network consensus}\label{sec:Sec4}
\subsection{Control of open inhomogeneous two-level quantum ensembles}
We consider an inhomogeneous open two-level ensemble with inhomogeneous parameter bound $E=0.2$. Members of the ensemble are governed by the following Hamiltonian:
\begin{equation}\label{eq:ensemble Hamiltonian}
      H(t)=\theta_0\frac{1}{2}\sigma_z+\theta_1u(t)(\sigma_x\cos\varphi+\sigma_y\sin\varphi),
\end{equation}
where $\varphi=0.8897$, $u(t)\in [-10, 10]$, and the Pauli operators are defined as:
\begin{equation}\label{eq:Pauli operators}
\sigma_{x}=\begin{pmatrix}
  0 & 1  \\
  1 & 0  \\
\end{pmatrix} , \ \ \ \
\sigma_{y}=\begin{pmatrix}
  0 & -i  \\
  i & 0  \\
\end{pmatrix} , \ \ \ \
\sigma_{z}=\begin{pmatrix}
  1 & 0  \\
  0 & -1  \\
\end{pmatrix}.
\end{equation}
For simplicity, we let the decoherence coefficients $\gamma_{k}=1$ and the Lindblad operators are given by \cite{Jirari and W. Potz 2005}
    	\begin{equation}\label{eq:Lindblad operators}
     	L_1=\left[ \begin{array}{cc}
     	0 & 0 \\
     	0.1 & 0
     	\end{array}\right],\
      	L_2=\left[ \begin{array}{cc}
     	0 & 0.2 \\
     	0 & 0
     	\end{array}\right],\
      	L_3=\left[ \begin{array}{cc}
     	0.2 & 0 \\
     	0 & 0
     	\end{array}\right],
     	\end{equation}\
where the items $L_1$ and $L_2$ correspond to relaxation and $L_3$ item characterizes the dephasing process. For a two-level quantum ensemble, $U_{1}, U_{2}$ and $U_{3}$ in (\ref{eq:density_operator}) can be chosen as $\sigma_{x}$, $\sigma_{y}$ and $\sigma_{z}$. The coherent vector for the density matrix is the Bloch vector
$$\textbf{r}=(x,y,z)^{T}=(\text{tr}(\sigma_{x}\rho), \text{tr}(\sigma_{y}\rho), \text{tr}(\sigma_{z}\rho)).$$
The dynamical equation for $\textbf{r}$ can be written as
\begin{equation}\label{Example1}
\begin{array}{ll}
  \dot{\textbf{r}}(t)
=&\left(
\begin{array}{ccc}
  -0.045 & -\theta_{0} & 0 \\
  \theta_{0} & -0.045  & 0 \\
  0 & 0 & -0.05 \\
\end{array}%
\right)\textbf{r}(t)+\left(%
\begin{array}{c}
  0\\
  0\\
  0.03\\
\end{array}%
\right)\\
&+\theta_{1}u(t)\left(
\begin{array}{ccc}
  0 & 0 & -2\sin\varphi \\
  0 & 0  & 2\cos\varphi \\
  2\sin\varphi & -2\cos\varphi & 0 \\
\end{array}%
\right)\textbf{r}(t).
\end{array}
\end{equation}
The average fitness function is given as
\begin{equation}\label{eq:average function for ensemble}
 E[J_{\varTheta}(u)]=\frac{1}{N}\sum_{\theta_0}\sum_{\theta_1} [1-\frac{1}{4}\parallel\textbf{r}_f -\textbf{r}_{\theta_0,\theta_1}(T)\parallel^2],
\end{equation}
where $N$ is the total number of the chosen samples. An upper bound of the fitness function is 1 although we do not a priori know the maximum that can be achieved. In the \emph{msMS}\_DE algorithm, we choose three samples for each parameter, and here we have $N=9$.
During learning control of the inhomogeneous quantum ensemble, we employ DE algorithms to seek the optimal control $u^*(t)$. Then, we apply the optimal control field to additional samples with inhomogeneous parameters $(\theta_0,\theta_1)$ following uniform distributions within $[0.8,1.2]$ to test its performance. We assume that the initial state $\rho_{0}$ and the target state $\rho_{f}$ are, respectively,
\begin{equation}
\rho_{0}=\begin{pmatrix}
  1 & 0  \\
  0 & 0  \\
\end{pmatrix} , \ \ \ \
\rho_{f}=\begin{pmatrix}
  0 & 0  \\
  0 & 1  \\
\end{pmatrix}.
\end{equation}
The target time $T=10$ and the time interval $[0,\ T]$ is equally divided into $D=200$ time steps, and $\Delta t=0.05$. The population size is set as $NP=50$ for all the algorithms in this example. The simulation is implemented on a MATLAB platform (version 8.3.0.532). The hardware environment for simulation is Intel(R)-Core(TM) i7-6700K CPU, dominant frequency @4.00GHz, and 16G(ARM).

\begin{figure}
\centering
\includegraphics[width=0.99\textwidth]{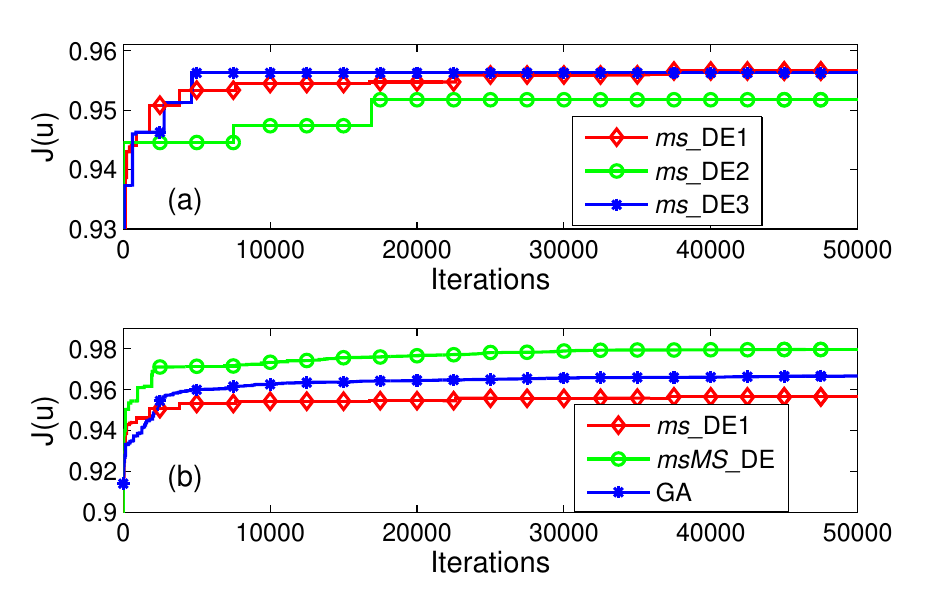}
\caption{(a) The training performance of the two-level open quantum ensemble via \emph{ms}\_DE1 ($F=0.9, CR=0.1$), \emph{ms}\_DE2 ($F=0.9, CR=0.9$) and \emph{ms}\_DE3 ($F=0.5, CR=0.3$). (b) The training performance of \emph{ms}\_DE1, GA and \emph{msMS}\_DE.}
\label{fig:ensemble DE}
\end{figure}

To demonstrate the performance of the proposed \emph{msMS}\_DE algorithm for the control problem of inhomogeneous quantum ensembles, we make performance comparison between it and \emph{ms}\_DE (DE with multiple samples, i.e., using the average fitness function of multiple samples) with various parameters. To begin with, we present the results for the traditional DE (i.e., ``DE/rand/1/bin") using multiple samples with three typical sets of control parameters. Three cases with different control parameters are labeled as ``\emph{ms}\_DE1" ($F=0.9, CR=0.1$), ``\emph{ms}\_DE2" ($F=0.9, CR=0.9$), and ``\emph{ms}\_DE3" ($F=0.5$, $CR=0.3$), and the training performance is presented in Fig. \ref{fig:ensemble DE}(a). It is clear that \emph{ms}\_DE1 and \emph{ms}\_DE3 have better performance than \emph{ms}\_DE2 for the quantum control problem. \emph{ms}\_DE1 can achieve the highest fitness $0.9566$ among these three cases. Note that the fitness has an upper bound 1 for this class of quantum control problems although its tight upper bound is less than 1. We then compare the training performance of \emph{ms}\_DE1, GA and \emph{msMS}\_DE, and the results are illustrated in Fig. \ref{fig:ensemble DE}(b). The \emph{msMS}\_DE algorithm achieves the highest fitness $J_{\max}= 0.9798$, while \emph{ms}\_DE1 and GA converge to a maximum value of 0.9566 and 0.9667, respectively. A comparison of testing performance for 2000 additional samples (i.e., 2000 quantum systems generated according to the inhomogeneous ensemble) and training times between DE1 (using one sample), \emph{ms}\_DE1, \emph{ms}\_DE2, \emph{ms}\_DE3, GA (with crossover probability $P_{c}=0.8$ and mutation probability $P_{m}=0.05$) and \emph{msMS}\_DE in Table 1 shows that \emph{msMS}\_DE is superior to \emph{ms}\_DE and DE1. More numerical results also show that \emph{msMS}\_DE usually can find the control field with the best robustness among these algorithms because \emph{msMS}\_DE employs mixed mutation strategies as well as average performance using multiple samples. \emph{ms}\_DE1, \emph{ms}\_DE2, \emph{ms}\_DE3, GA and \emph{msMS}\_DE also take similar time to find an optimal solution for the ensemble control problem. For example, \emph{msMS}\_DE takes 9 hours 20 minutes and GA takes 10 hours 18 minutes 14 seconds.

\begin{table}[!htp]
\scalebox{0.85}{
\begin{tabular}{|c|c|c|c|}
\hline
Algorithm & parameters & training time & $\bar{\textbf{J}}(u)$ \\
\hline
DE1 & $CR=0.1, F=0.9, N=1$ & 1h10m47s & 0.9408\\
\hline
\emph{ms}\_DE1 & $CR=0.1, F=0.9, N=9$ & 9h27m47s & 0.9610\\
\hline
\emph{ms}\_DE2 & $CR=0.9, F=0.9, N=9$ & 9h20m15s & 0.9537\\
\hline
\emph{ms}\_DE3 & $CR=0.3, F=0.5, N=9$ & 9h40m5s & 0.9601 \\
\hline
GA & $P_{c}=0.8$, $P_{m}=0.05$, $N=9$ & 10h18m14s & 0.9691\\
\hline
\emph{msMS}\_DE & $F=\text{N}(0.5,0.3)$, & 9h20m0s & 0.9803\\
& $CR=\text{N}(0.5,0.1), N=9$ & & \\
\hline
\end{tabular}}
\caption{Performance comparison of different algorithms}
\label{tab:DE variants}
\end{table}

\subsection{Consensus in superconducting qubit networks}\label{sec:Sec6}
The nodes in a quantum network could be photons, electrons, or other quantum systems. In this section, we consider a quantum network that consists of superconducting qubits as its nodes. Superconducting quantum circuits based on Josephson junctions are one promising candidate for building the hardware of quantum computers \cite{You and Nori 2005}.
Superconducting qubits have been widely investigated theoretically as well as implemented experimentally since they could be easily embedded in nanometer-scale electronic devices and scaled up to provide a large number of qubits for quantum computation \cite{Makhlin et al 1999}. One may manipulate superconducting qubits by adjusting external parameters such as voltages and currents or by tuning the coupling between two superconducting qubits \cite{Bialczak et al 2011}.

A typical class of superconducting qubits is charge qubits when $E_C\gg E_J$ where $E_C$ denotes the charging energy and $E_J$ denotes the Josephson coupling energy \cite{You et al 2003}.
The equivalent Hamiltonian of a charge qubit can be described as \cite{Dong 2015SR}, \cite{Dong 2016SR}
      	\begin{equation}\label{eq:reduced Hamiltonian of one qubit}
        H=F_z(V_g)\sigma_z-F_x(\varPhi)\sigma_x,
     	\end{equation}
where $F_z(V_g)$ can be adjusted through an external voltage $V_g$, and $F_x(\varPhi)$ corresponds to a tunable effective coupling with the external magnetic flux $\varPhi$ in the superconducting quantum interference device. Hence, $F_z(V_g)$ and $F_x(\varPhi)$ are related to external control fields.

Now, consider a quantum network consisting of three superconducting qubits with control fields acting on all qubits. We denote $\sigma_x^{(12)}=\sigma_x\otimes \sigma_x \otimes I $, $\sigma_x^{(23)}= I\otimes \sigma_x \otimes \sigma_x$, $\sigma_x^{(13)}= \sigma_x \otimes I \otimes \sigma_x $. Its free Hamiltonian can be described as
\begin{equation}\label{eq:free Hamiltonians of networks}
       H_0=\omega_{12}\sigma_x^{(12)}+\omega_{23}\sigma_x^{(23)}+\omega_{13}\sigma_x^{(13)}.
\end{equation}
Denote $\sigma_x^{(1)}=\sigma_x\otimes I \otimes I $, $\sigma_x^{(2)}= I \otimes \sigma_x \otimes I$, $\sigma_x^{(3)}= I \otimes I \otimes \sigma_x$, and $\sigma_z^{(1)}=\sigma_z\otimes I \otimes I $, $\sigma_z^{(2)}= I \otimes \sigma_z \otimes I$, $\sigma_z^{(3)}= I \otimes I \otimes \sigma_z$. We have the control Hamiltonian in the following form
\begin{equation}\label{eq:control Hamiltonians of networks}
       H_u(t)=\ u^{x}_{1}\sigma_x^{(1)}+u^{z}_{1}\sigma_z^{(1)}+u^{x}_{2}\sigma_x^{(2)}+u^{z}_2\sigma_z^{(2)}+u^{x}_3\sigma_x^{(3)}+u^{z}_{3}\sigma_z^{(3)}.
\end{equation}

The goal is to drive the quantum network from an arbitrary initial state (usually three qubits having different reduced states) to a consensus state. Furthermore, if we withdraw the external control fields, the quantum network will remain in the consensus state under the free Hamiltonian. Denote $\mathbf{1}_{n}$ as an $n$-dimensional matrix with all of its elements being $1$. Let the target state be $\bar{\rho}=\frac{1}{8}\mathbf{1}_{8}$. We have the following result.
\begin{proposition}
The state $\bar{\rho}=\frac{1}{8}\mathbf{1}_{8}$ is a consensus state for the three qubit network. Also, $\bar{\rho}$ is invariant under the action of free Hamiltonian $H_0=\omega_{12}\sigma_x^{(12)}+\omega_{23}\sigma_x^{(23)}+\omega_{13}\sigma_x^{(13)}$.
\end{proposition}

\textbf{Proof}\\
For $\bar{\rho}=\frac{1}{8}\mathbf{1}_{8}$, we can calculate the reduced states for three nodes as follows:
$$\bar{\rho}_{1}=\frac{1}{2}\mathbf{1}_{2},\ \bar{\rho}_{2}=\frac{1}{2}\mathbf{1}_{2},\ \bar{\rho}_{3}=\frac{1}{2}\mathbf{1}_{2}.$$
It is clear that $\bar{\rho}_{1}=\bar{\rho}_{2}=\bar{\rho}_{3}$. That is, the state $\bar{\rho}$ is a consensus state for the three qubit network according to Definition 2.

A direct calculation shows that $[H_{0}, \bar{\rho}]=0$. Hence, $$\dot{\bar{\rho}}=[H_{0}, \bar{\rho}]=0.$$ That is, $\bar{\rho}$ is invariant under the action of free Hamiltonian $H_0$.

\begin{figure}
\centering
\includegraphics[width=1.0\textwidth]{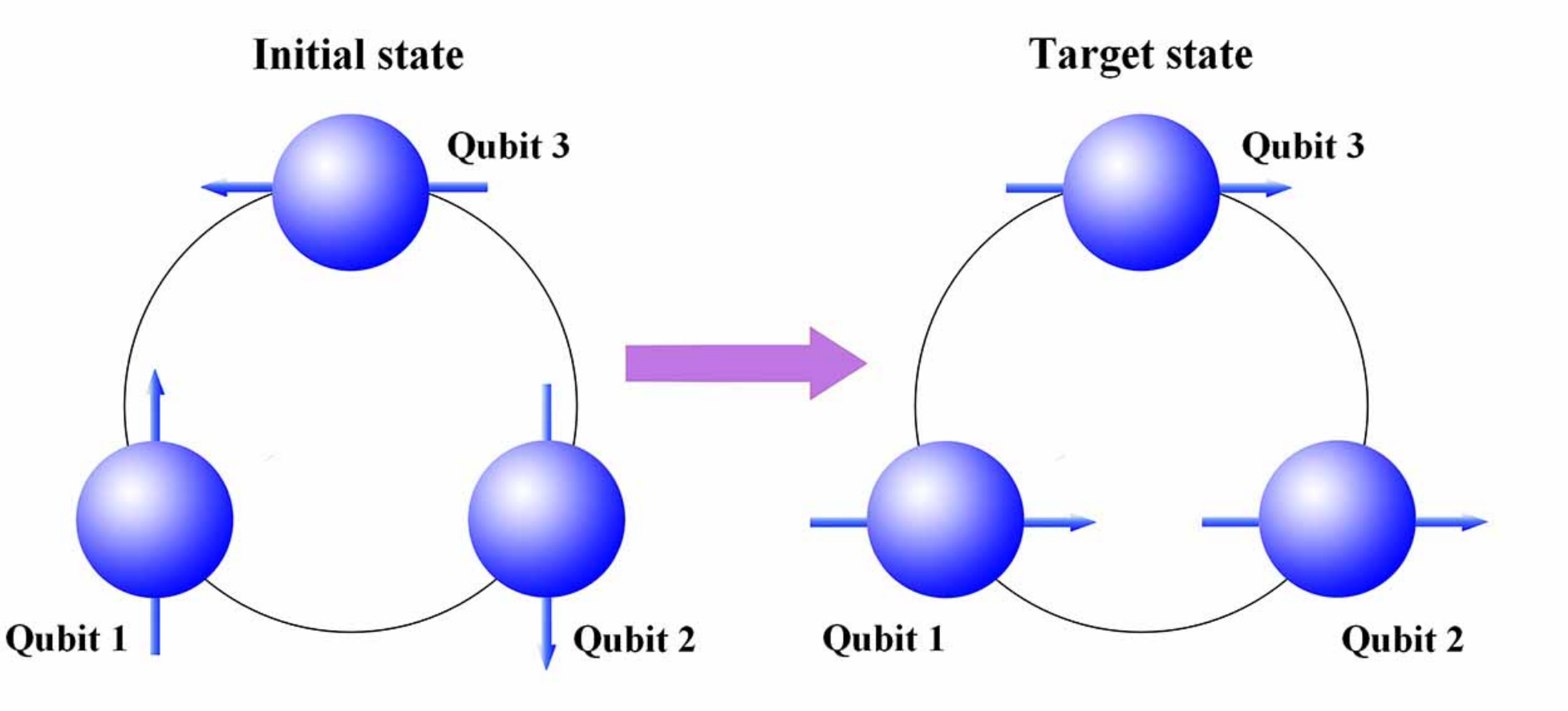}
\caption{The initial and target (reduced) states of the three qubits on the Bloch sphere.}
\label{fig:qubit network state transfer}
\end{figure}

The initial state is set as $\bar{\rho}^{0}=\rho^{0}_{1}\otimes \rho^{0}_{2}\otimes \rho^{0}_{3}$ where
$$\rho^{0}_{1}=\left(
                 \begin{array}{cc}
                   1 & 0 \\
                   0 & 0 \\
                 \end{array}
               \right), \
               \rho^{0}_{2}=\left(
                 \begin{array}{cc}
                   \frac{1}{2} & -\frac{1}{2} \\
                   -\frac{1}{2} & \frac{1}{2} \\
                 \end{array}
               \right), \ \rho^{0}_{3}=\left(
                 \begin{array}{cc}
                   0 & 0 \\
                   0 & 1 \\
                 \end{array}
               \right).
$$
The initial and target states of qubit network are illustrated in Fig. \ref{fig:qubit network state transfer}.

\begin{figure}
\centering
\includegraphics[width=1.0\textwidth]{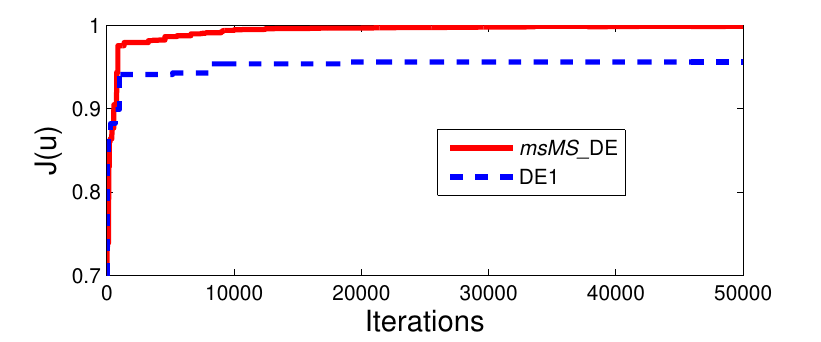}
\caption{The training performance of superconducting qubit network via DE1 and \emph{msMS}\_DE.}
\label{fig:qubit network training}
\end{figure}

\begin{figure}
\centering
\includegraphics[width=1.0\textwidth]{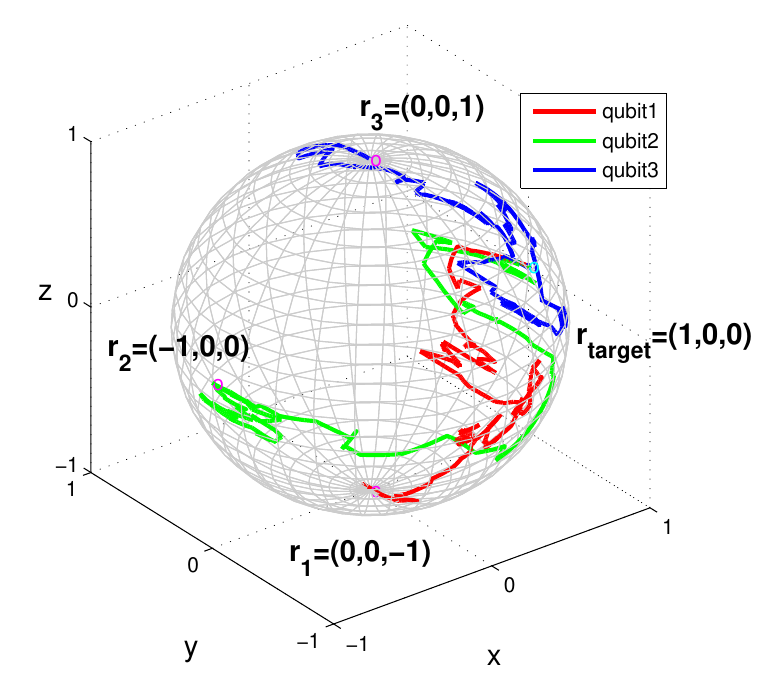}
\caption{The reduced states of three qubits asymptotically converge to the same trajectory for $\theta_x=\theta_z=1$ using the control learned by \emph{msMS}\_DE.}
\label{fig:qubit network orbits}
\end{figure}

\begin{figure}
\centering
\includegraphics[width=1.0\textwidth]{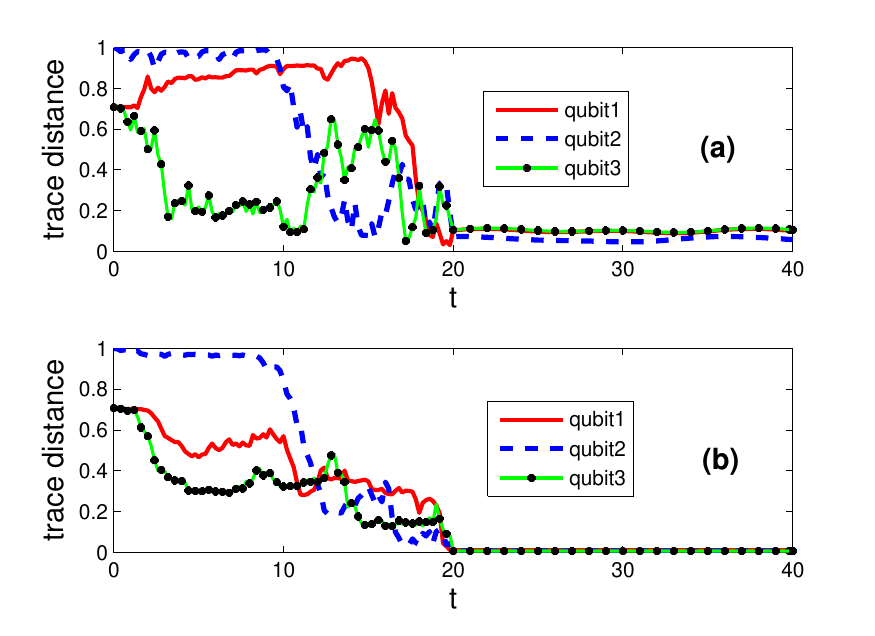}
\caption{The testing performance of the qubit network. (a) Average evolution curves of trace distances  of three qubits for 2000 samples using the control field learned by DE1. (b) Average evolution curves of trace distances  of three qubits for 2000 samples using the control field learned by \emph{msMS}\_DE.}
\label{fig:distance1}
\end{figure}

\begin{figure}
\centering
\includegraphics[width=1.0\textwidth]{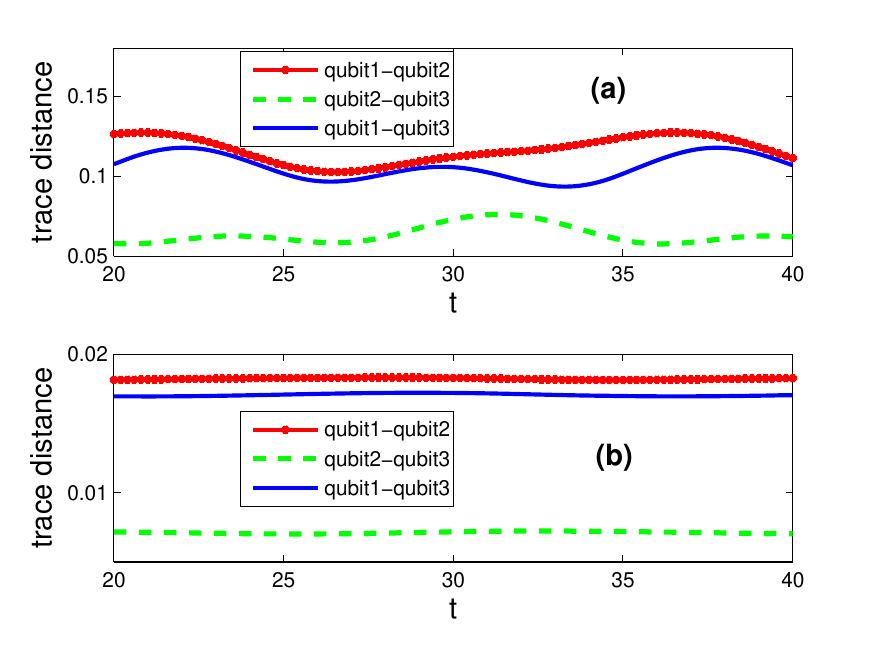}
\caption{The trace distances between different qubits in the qubit network with free Hamiltonian. (a) Average evolution curves of trace distances between the three qubits for 2000 samples (DE1). (b) Average evolution curves of trace distances  between the three qubits for 2000 samples (\emph{msMS}\_DE).}
\label{fig:distance2}
\end{figure}

In practical applications, there may exist variations in magnetic fields and electric fields in superconducting qubits. The practical control Hamiltonian is assumed to be
\begin{equation}\label{eq:control Hamiltonians of networks}
\begin{aligned}
       H_u(t)=&\theta_{x}u^{x}_{1}\sigma_x^{(1)}+\theta_{z}u^{z}_{1}\sigma_z^{(1)}+\theta_{x}u^{x}_{2}\sigma_x^{(2)}+\theta_{z}u^{z}_2\sigma_z^{(2)}\\
       &+\theta_{x}u^{x}_3\sigma_x^{(3)}+\theta_{z}u^{z}_{3}\sigma_z^{(3)}.
\end{aligned}
\end{equation}
We apply the \emph{msMS}\_DE algorithm to search for a robust control field to reach a consensus state in the above quantum network. The stimulation parameters are set as: the population size $NP=100$, the time internal $[0,20]$ ns is equally divided into 100 smaller time steps (i.e., $D=100$), the control field components are $u^{x}_{1}$, $u^{z}_{1}$, $u^{x}_{2}$, $u^{z}_{2}$, $u^{x}_{3}$, $u^{z}_{3}$ $\in[0,1]$ GHz. Considering that we can manipulate a superconducting circuit at the nanosecond scale and the coupling between two superconducting qubits can be at the scale of $100 \ \text{MHz}$ \cite{You and Nori 2005}, \cite{Dong 2015SR}, let $\omega_{12}=\omega_{23}=\omega_{13}=0.1\ \text{GHz}$.
We assume that $\theta_x\in [0.98, 1.02]$ and $\theta_z\in [0.98, 1.02]$ (i.e., $E=0.02$). For each uncertain parameter, we choose three samples and have $N=9$ samples for training. We employ DE1 (one sample) for comparison. The training performance of driving qubit network is illustrated in Fig. \ref{fig:qubit network training}. As we can see, \emph{msMS}\_DE achieves a rather high fitness $J_{\max}=0.9988$ (where 1 is an upper bound of $J$), while DE1 achieves the fitness of $J_{\max}=0.9561$. For the case $\theta_x=\theta_z=1$, Fig. \ref{fig:qubit network orbits} shows the reduced states of three qubits from different trajectories asymptotically converging to the same trajectory using the control learned from \emph{msMS}\_DE. Based on the control fields from DE1 and \emph{msMS}\_ED, we test 2000 additional samples using the trace distance defined as (for $i=1,2,3$) $$||\rho_{i}-\frac{1}{2}\mathbf{1}_{2}||_{\text{Tr}}=\frac{1}{2}\text{Tr}\sqrt{(\rho_{i}-\frac{1}{2}\mathbf{1}_{2})^{\dagger}(\rho_{i}-\frac{1}{2}\mathbf{1}_{2})}=\frac{1}{2}\sum_{j=1}^{2}|\lambda_{j}|$$
where $\lambda_{j}$ are eigenvalues of $\rho_{i}-\frac{1}{2}\mathbf{1}_{2}$. Since the maximum trace distance between two quantum states may be 1, we define the relative error between two quantum states $\rho_{i}$ and $\rho_{j}$ as $||\rho_{i}-\rho_{j}||_{\text{Tr}}\times 100\%$. Fig. \ref{fig:distance1} shows that the relative error between each qubit and its target state always remains below $1.2\%$ for the case using \emph{msMS}\_DE while the relative error between each qubit and its target state may exceed $10.0\%$ for the case using DE1. The trace distance between all the three qubits can approximately reach 0 in Fig. \ref{fig:distance1}(b) while the distance between these qubits in Fig. \ref{fig:distance1}(a) cannot converge to 0. We further show the trace distances between the quantum states of different qubits after the control Hamiltonian is withdrawn in Fig. \ref{fig:distance2}. It is clear that the relative errors between the reduced states are always below $2.0\%$ for the case of \emph{msMS}\_DE while the relative errors may exceed $12.0\%$ for the case of DE1. The results demonstrate that the approximate consensus state achieved using \emph{msMS}\_DE has much better stability than that obtained using DE1.


%

\section{Experimental results on femtosecond laser control systems}\label{sec:Sec5}
The following quantum control experiments were carried out in the Department of Chemistry at Princeton University.

\subsection{Experimental setup}
The experimental setup contains three major components: 1) a fs laser system, 2) a pulse shaper and 3) a time-of-flight mass spectrometry (TOF-MS).  Briefly, the fs laser system (KMlab, Dragon) consists of a Ti:sapphire oscillator and a amplifier, which produces 1 mJ, 25 fs pulses centered at 790 nm. The laser pulses are introduced into a pulse shaper with a programmable dual-mask liquid crystal spatial light modulator (SLM). The SLM has the capability of independent phase and amplitude modulation and has 640 pixels with 0.2 nm/pixel resolution \cite{Tibbetts JPC}, \cite{Tibbetts PCCP}. Every 8 adjacent pixels are bundled together to form an array of 80 ``grouped pixels", which are the control variables. Each control variable can have a phase value between 0 and $2\pi$, and an amplitude value between 0 and 1.  In this experiment, we do phase-only control, with all the amplitude values fixed at 1. The shaped laser pulses out of the shaper are focused into a vacuum chamber, where photoionization and photofragmentation occurs. The fragment ions are separated with a set of ion lens and passing through a TOF tube before being collected with a micro-channel plate detector. The mass spectrometry signals are recorded with a fast oscilloscope, which accumulates 3000 laser shots in one second before sending the average signal to a personal computer for further analysis. A small fraction of the beam ($<5\%$) is separated from the main beam and focused into a GaP photodiode (Thorlab, DET25K), which collects signals arising from TPA.

\subsection{Optimization of TPA signal}
\begin{figure}
\centering
\includegraphics[width=0.9\textwidth]{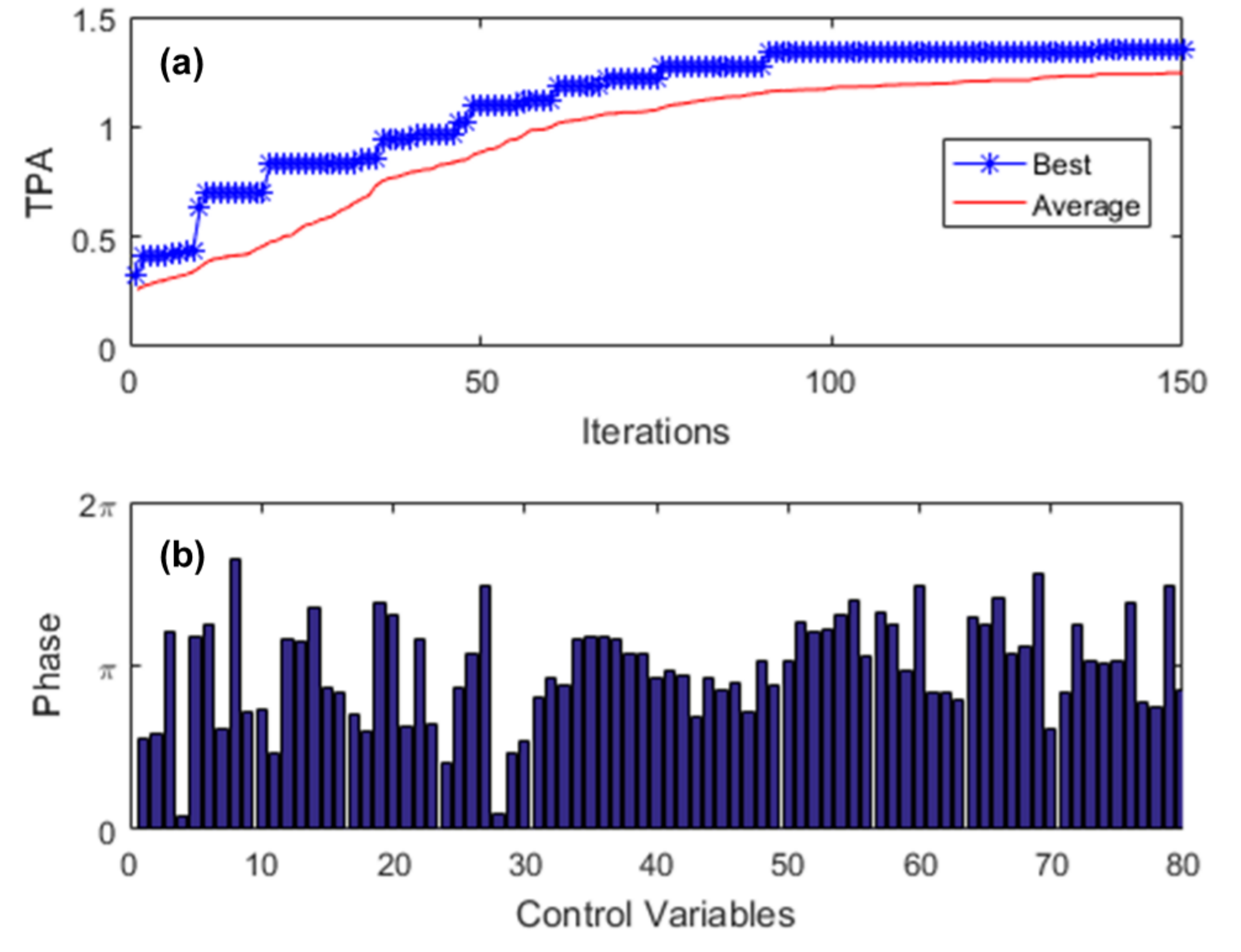}
\caption{Experimental result for optimizing the TPA signal using the \emph{msMS}\_DE algorithm. (a) TPA signal vs iterations, where `Best' represents the maximum fitness and `Average' represents the average fitness of all individuals during each iteration. (b) Optimized phases of 80 control variables for the final optimal result corresponding to the maximum fitness.}
\label{Experiment_TPA}
\end{figure}

A preliminary task is to optimize the TPA signal, which is a convenient way to identify the shortest pulse that removes the residual high-order dispersion in the amplifier output. The parameter setting is as follows: $D = 80$, $NP = 30$ and $N = 3$. The control variables are the phases. The fitness function corresponds to the TPA signal and an average fitness of three samples is used. Three samples for each individual are selected as follows: The first sample comes from the current individual, denoted as $X_{i}^{1}=[x_{i}^{1}, x_{i}^{2}, \dots, x_{i}^{80}]^{T}$, the second sample is generated by adding a random fluctuation between 0 and $0.1\pi$ to each component of the current individual, i.e., $X_{i}^{2}=[x_{i}^{1}+0.05\text{rand}(0, 1)\times 2\pi, x_{i}^{2}+0.05\text{rand}(0, 1)\times 2\pi, \dots, x_{i}^{80}+0.05\text{rand}(0, 1)\times 2\pi]^{T}$, and the third sample is selected as $X_{i}^{3}=[x_{i}^{1}-0.05\text{rand}(0, 1)\times 2\pi, x_{i}^{2}-0.05\text{rand}(0, 1)\times 2\pi, \dots, x_{i}^{80}-0.05\text{rand}(0, 1)\times 2\pi]^{T}$. This means that each control variable is permitted to have up to $5\%$ (of the maximum phase) additive noise. The interaction between the algorithm and the modulation of SLM is accomplished by LabVIEW software. An experimentally reasonable termination condition of $150$ generations (iterations) is used. For $150$ iterations, it approximately takes five and a half hours to run the experiment. For each generation, a total of 90,000 signal measurements were made. The experimental result is shown in Fig. \ref{Experiment_TPA} where TPA signal (measured using a GaP fast photodiode) during each iteration is presented in Fig. \ref{Experiment_TPA}(a) and the optimized phases of 80 control variables for the final optimal result is given in Fig. \ref{Experiment_TPA}(b). After 150 generations, the best average TPA signal for three samples can reach 1.35.

\subsection{Fragmentation control}
We consider the fragmentation control of $\text{CH}_2\text{BrI}$, where the fitness is defined as maximization of the photofragment ratio of $\text{CH}_2\text{Br}^{+}/\text{CH}_2\text{I}^{+}$, while the control variables are the phases. The parameter setting is the same as that in Section V.B: $D = 80$ and $NP = 30$.  DE algorithms were employed to optimize the phases of 80 control variables. Here, we apply both DE1 and \emph{msMS}\_DE for comparison.

\begin{figure}
\centering
\includegraphics[width=0.9\textwidth]{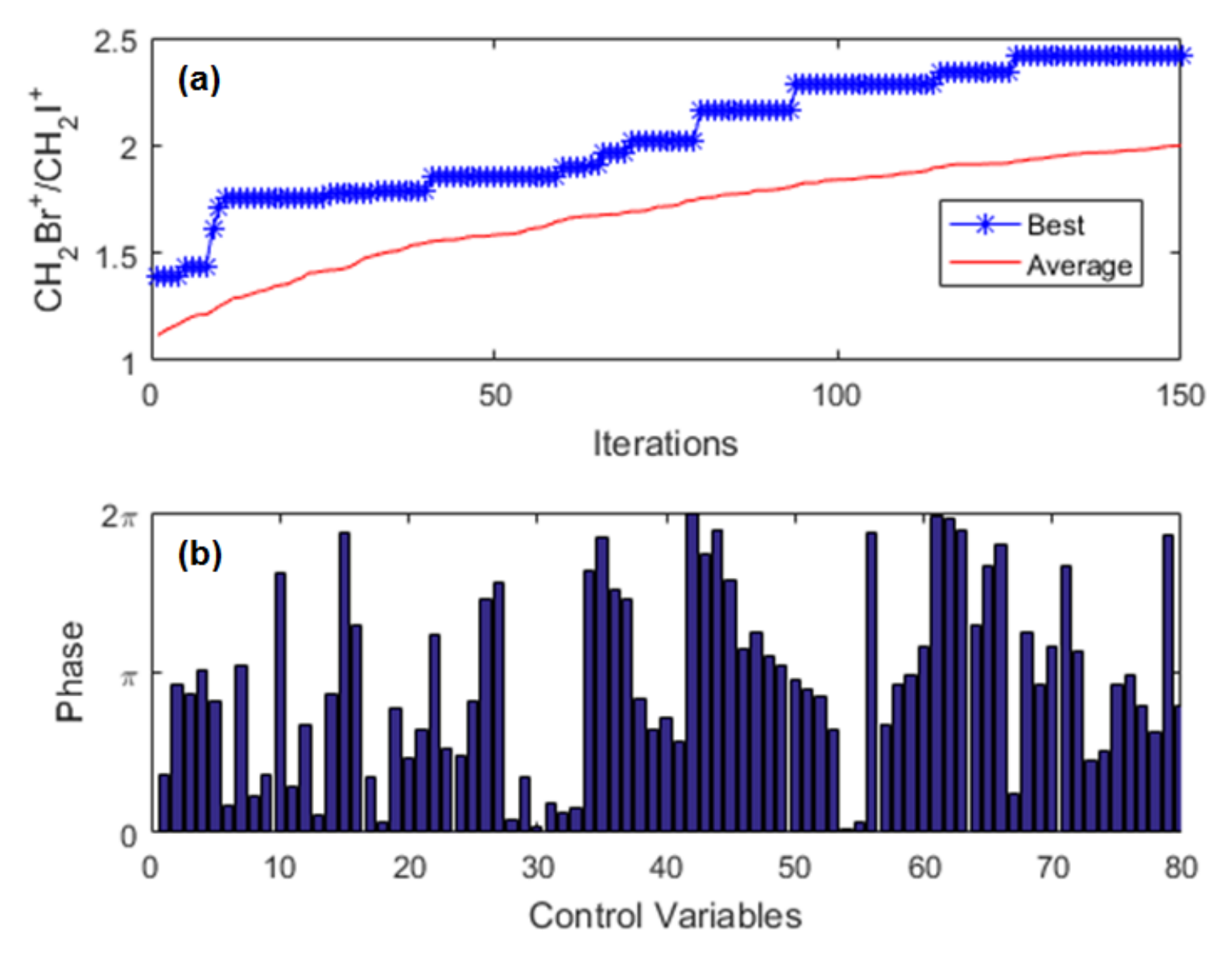}
\caption{Experimental result for optimizing the ratio between the products $\text{CH}_2\text{Br}^{+}$ and $\text{CH}_2\text{I}^{+}$ using DE1. (a) Ratio $\text{CH}_2\text{Br}^{+}/\text{CH}_2\text{I}^{+}$ vs iterations, where `Best' represents the maximum fitness and `Average' represents the average fitness of all individuals during each iteration. (b) Optimized phases of 80 control variables for the final optimal result corresponding to the maximum fitness.}
\label{Experiment_DE1}
\end{figure}

Figure \ref{Experiment_DE1} shows the experimental results using the DE1 algorithm, where the ratio $\text{CH}_2\text{Br}^{+}/\text{CH}_2\text{I}^{+}$ as the fitness function is presented in Fig. \ref{Experiment_DE1}(a) and the optimized phases of 80 control variables for the final optimal result is given in Fig. \ref{Experiment_DE1}(b). In Fig. \ref{Experiment_DE1}(a), `Best' represents the maximum fitness and `Average' represents the average fitness of all individuals during each iteration. With 150 iterations, DE1 can find an optimized pulse to make $\text{CH}_2\text{Br}^{+}/\text{CH}_2\text{I}^{+}$ to achieve 2.41.

\begin{figure}
\centering
\includegraphics[width=0.9\textwidth]{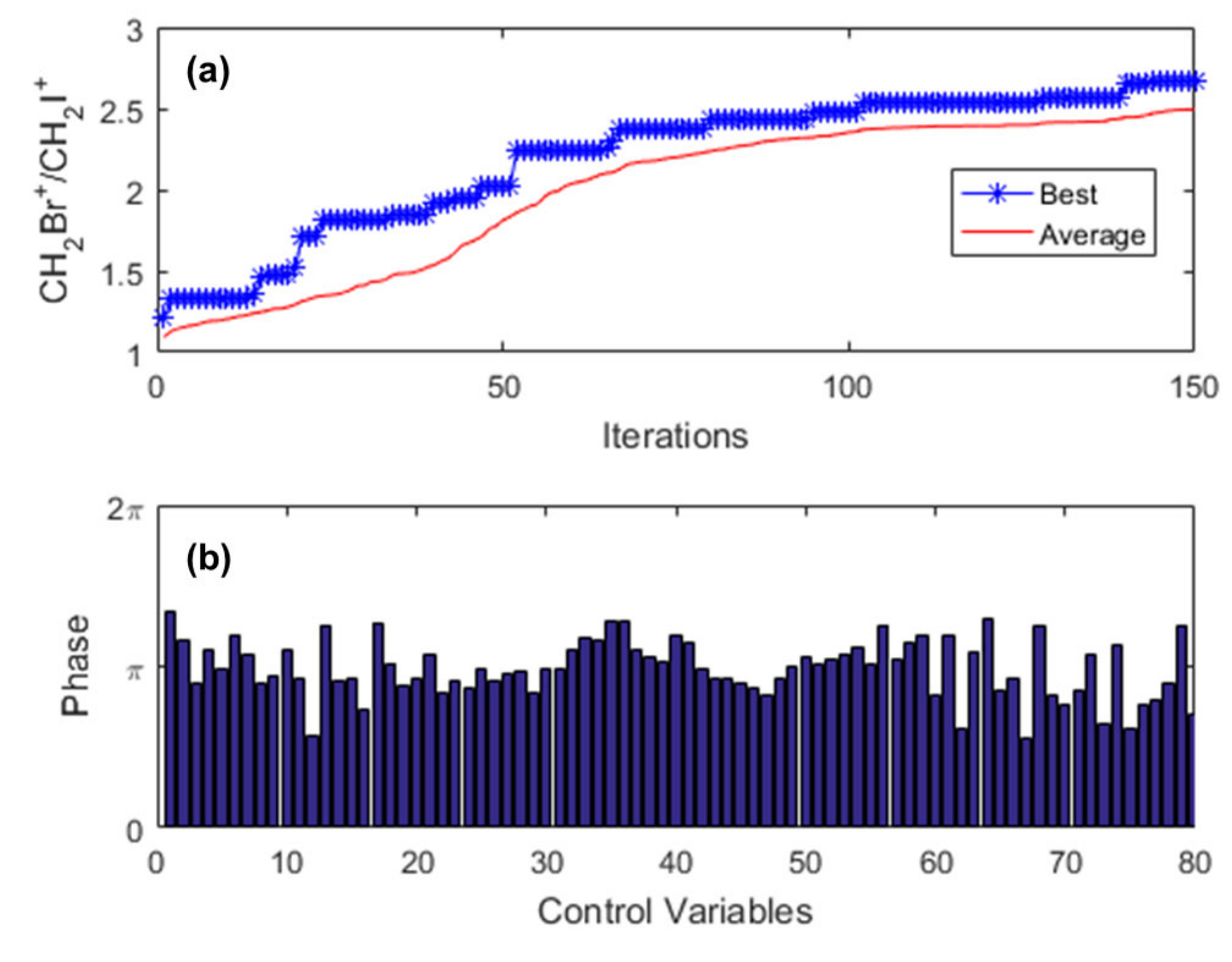}
\caption{Experimental result for optimizing the ratio between the products $\text{CH}_2\text{Br}^{+}$ and $\text{CH}_2\text{I}^{+}$ using \emph{msMS}\_DE. (a) Ratio $\text{CH}_2\text{Br}^{+}/\text{CH}_2\text{I}^{+}$ vs iterations, where `Best' represents the maximum fitness and `Average' represents the average fitness of all individuals during each iteration. (b) Optimized phases of 80 control variables for the final optimal result corresponding to the maximum fitness.}
\label{Experiment_msMS_DE}
\end{figure}

Figure \ref{Experiment_msMS_DE} shows the results from the \emph{msMS}\_DE algorithm, in which three samples are measured in each experiment. Three samples for each individual were selected using the same method as that in the experiments of optimizing TPA signals. With 150 iterations, \emph{msMS}\_DE can find an optimal pulse for making the average $\text{CH}_2\text{Br}^{+}/\text{CH}_2\text{I}^{+}$ of three samples to achieve 2.67. The experimental results are shown in Fig. \ref{Experiment_msMS_DE}, where the average ratio $\text{CH}_2\text{Br}^{+}/\text{CH}_2\text{I}^{+}$ of three samples as the fitness function is presented in Fig. \ref{Experiment_msMS_DE}(a) and the optimized phases of 80 control variables for the final optimal result are given in Fig. \ref{Experiment_msMS_DE}(b).

\begin{figure}
\centering
\includegraphics[width=0.9\textwidth]{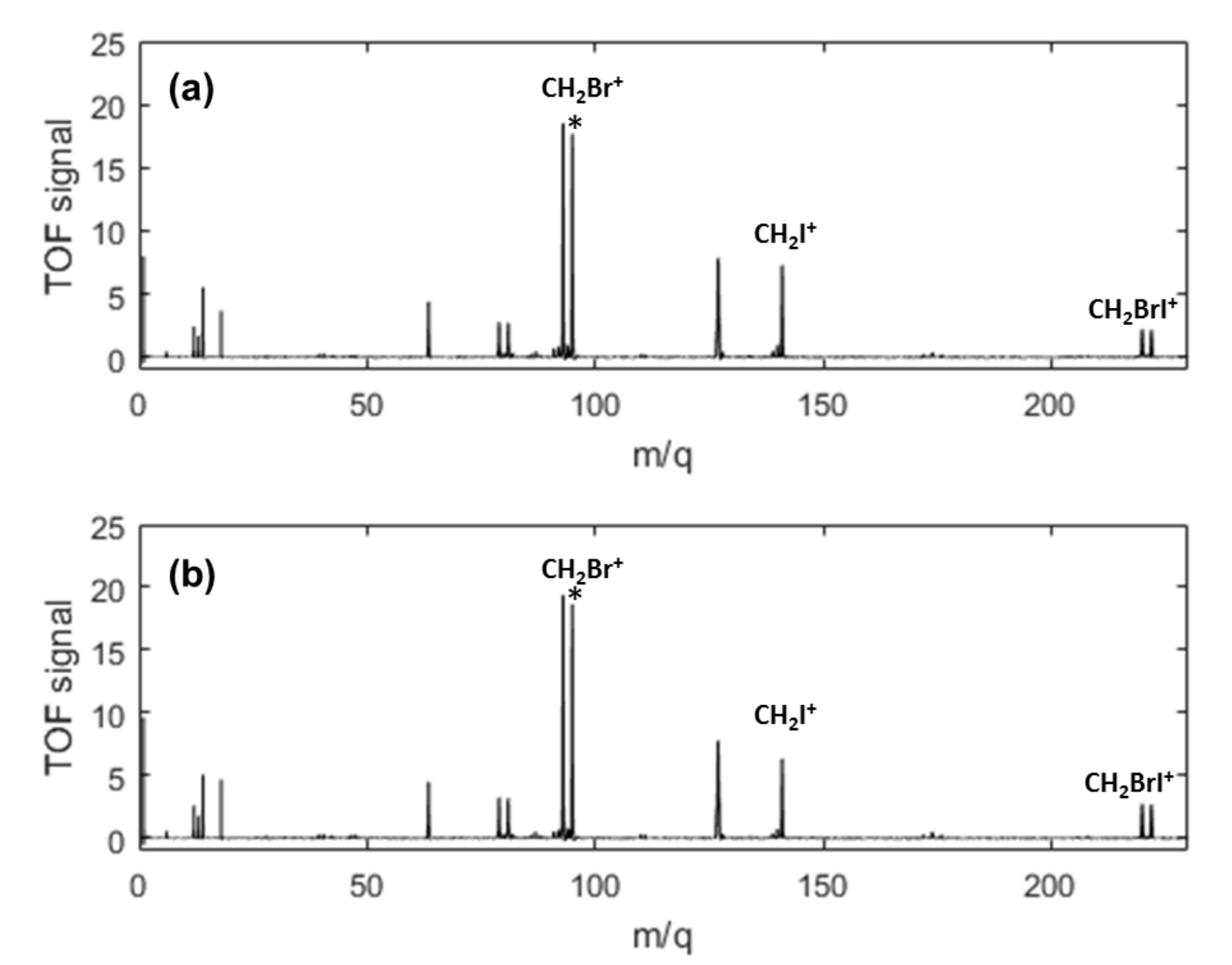}
\caption{Experimental results for TOF signal where $m$ represents mass and $q$ represents charge. (a) The average TOF signal of 100 testing results with random noises between $-7.5\% $ and $+7.5\% $ for the femtosecond pulse optimized by DE1. (b) The average TOF signal of 100 testing results with random noises between $-7.5\% $ and $+7.5\% $ for the femtosecond pulse optimized by \emph{msMS}\_DE.}
\label{Experiment_TOF}
\end{figure}

After we obtained the optimal femtosecond control pulses using DE1 and \emph{msMS}\_DE, we can test the performance of the optimal pulses. The testing results are shown in Fig. \ref{Experiment_TOF}, where Fig. \ref{Experiment_TOF}(a) is the average TOF signal of 100 testing results with random noises between $-7.5\% $ and $+7.5\% $ (with respect to the maximum phase $2\pi$) for the femtosecond pulse optimized by DE1. In other words, if we denote the best individual as $X_{\text{best}}=[x_{b}^{1}, x_{b}^{2}, \dots, x_{b}^{80}]^{T}$, these 100 testing samples can be written as $X_{s}^{k}=[x_{b}^{1}+0.075(2\text{rand}(0, 1)-1)\times 2\pi, x_{b}^{2}+0.075(2\text{rand}(0, 1)-1)\times 2\pi, \dots, x_{b}^{80}+0.075(2\text{rand}(0, 1)-1)\times 2\pi]^{T}$.
Fig. \ref{Experiment_TOF}(b) shows the average TOF signal of 100 testing results for the femtosecond pulse optimized by \emph{msMS}\_DE. The average $\text{CH}_2\text{Br}^{+}/\text{CH}_2\text{I}^{+}$ of the 100 testing samples can achieve 2.61 for the pulse from \emph{msMS}\_DE while the average $\text{CH}_2\text{Br}^{+}/\text{CH}_2\text{I}^{+}$ is only 2.12 for the pulse from DE1. It is clearly evident that \emph{msMS}\_DE outperforms DE1 in terms of reaching a better objective fitness value in the presence of phase noise (e.g., shot-to-shot variations in reproducing laser source).

\section{CONCLUSION}\label{sec:Sec7}
In order to solve three classes of quantum robust control problems, we have proposed an improved \emph{msMS}\_DE algorithm using multiple samples for fitness evaluation and a mixed strategy for mutation. The \emph{msMS}\_DE algorithm shows excellent performance for the control problem of open inhomogeneous quantum ensembles and the consensus problem of quantum networks with uncertainties. We have experimentally implemented \emph{msMS}\_DE on femtosecond laser control in the laboratory to generate good TPA signal and control fragmentation of $\text{CH}_2\text{BrI}$.
In future research, there is room for exploring the use of DE for emerging quantum control engineering. For example, it is worth presenting a comprehensive comparative investigation on the performance of different evolutionary computation algorithms and their variants for optimal control of molecules using femtosecond laser pulses, and exploring efficient algorithms for laboratory applications with constraints. More efficient DE algorithms need to be developed for high-dimensional quantum control problems \cite{Palittapongarnpim et al 2017}.

\appendix
\renewcommand\thesection{\appendixname~\Alph{section}}
\renewcommand\theequation{\Alph{section}.\arabic{equation}}

\section{DE Algorithm}\label{subsec:basic DE}
In DE, the individual trial solutions are termed as \emph{parameter vectors} or \emph{genomes}, usually represented in a vector $ X=[x_1, x_2, \cdots, x_D ]^T$ where each parameter $x_i$ is a real number. Solving an optimization problem using DE is a search for a parameter vector to minimize or maximize a fitness function (objective function) $f(X)$. The following operations are used to evolve a population of  $D$-dimensional parameter vectors until a ``best" individual is generated and found.

(a) \textbf{Initialization}. DE searches for a global optimum point in a $D$-dimensional real parameter space $\Re^D$. Here, we denote the population at the current generation as $X_{i, G}=({x_{i, G}^1, \cdots, x_{i, G}^D})$, $i=1, ..., NP$ and let  $x_{i,G}^j \in[x_{\min}^j,x_{\max}^j]$, $(j=1,2,...,D)$, since these parameters correspond to physical variables with relevant bounds. We generally initialize the population (at $G=0$) as follows \cite{Mallipeddi et 2011}:
 \begin{equation}\label{eq:initialization}
  x_{i,0}^j=x_{\min}^j+\textup{rand}(0,1)\cdot(x_{\max}^j-x_{\min}^j), \quad j=1,2,...,D,
 \end{equation}
where $\textup{rand}(0,1)$ is a uniformly distributed random number.

(b) \textbf{Mutation}. In DE, the key to ``mutation" is to generate a difference vector by choosing three other distinct parameter vectors from the current generation (say, $X_{r_1}$, $X_{r_2}$, $X_{r_3}$). The indices $r_1, r_2, r_3 \in \{1, ..., NP\}$ are mutually exclusive integers randomly generated within the range $[1,NP]$ and $r_1, r_2, r_3\neq i$. The donor vector $V_{i,G+1}$ are generated by
    \begin{equation}\label{eq:mutation}
    V_{i, G}=X_{r_1, G}+F \cdot(X_{r_2,G}-X_{r_3, G}),
    \end{equation}
    where $F$ is a positive control parameter having a typical value in the interval $[0.4,1]$.

(c) \textbf{Crossover}. A crossover operation comes into play after the mutation to enhance the potential diversity of the population. The DE family has two types of crossover operations (i.e., binomial and exponential). The binomial (uniform) crossover is described as
\begin{equation}\label{eq:crossover}
    u_{i, G}^j=\left \{
\begin{split} \displaystyle
    v_{i, G}^j, \ \ \ \text{if}\ \text{rand}(j) \leq CR \ \text{or} \ j=\textup{rand}(1, D), \\
    x_{i, G}^j, \ \ \text{if} \ \text{rand}(j)>CR\ \text{and} \ j \neq \textup{rand}(1, D),\\
\end{split}\right.
\end{equation}
where $j=1, 2, ..., D$ and $\text{rand}(j)\in[0, 1]$ is a uniform random number.
$CR$ is a given constant within the range $[0,1)$, $\text{rand}(1, D)\in \{1, 2, ..., D\}$ is a randomly chosen index.

(d) \textbf{Selection}. To keep the population size constant over subsequent generations, DE uses the following selection operation to determine whether the target vector or the trial vector survives to the next generation:
    \begin{equation}\label{eq:selection}
    X_{i, G+1}=\left \{
\begin{split} \displaystyle
    U_{i, G}, \ \text{if} \quad f(U_{i, G})\geq f(X_{i, G}), \\
    X_{i, G}, \ \text{otherwise}. \ \ \ \ \ \ \ \ \ \ \ \ \ \
\end{split}\right.
    \end{equation}
If the new trial vector yields an equal or lower value of the objective function (assuming that minimization of the objective function is the goal), it replaces the corresponding target vector in the next generation; otherwise the target vector survives.

Usually, it is the mutation operation that demarcates one DE scheme from another. The DE strategy with the mutation in (\ref{eq:mutation}) is referred to as ``DE/rand/1" using the notation ``DE/x/y", where $x$ represents a string denoting the base vector to be perturbed, $y$ is the number of difference vectors considered for perturbation of $x$. Furthermore, when crossover is also considered,  the notation ``DE/x/y/z" is used, where $z$ stands for the type of crossover (bin: binomial, exp: exponential). DE variants with different mutation strategies usually have different performance for solving optimization problems.
Regarding control parameter $F$, Storn and Price \cite{Storn and Price 1995} have proposed that a good initial choice was 0.5 and the range of $F$ is usually set $[0.4,1]$. The crossover rate $CR$ may be $CR\in[0,1]$. Several results also proposed the techniques of self-adaptation to automatically find an optimal set of control parameters \cite{Das and Suganthan 2011}, \cite{Qin et al 2009} to provide improved performance.


\end{document}